\documentclass[aps,prl,twocolumn,superscriptaddress,preprintnumbers,showpacs]{revtex4}
\usepackage{graphics}
\def\eq#1\en{\begin{equation} #1 \end{equation}}

\def\eqa#1\ena{\begin{eqnarray} #1 \end{eqnarray}}

\usepackage{graphicx,color}

\begin{document}


\title{Dark consequences from light neutrino condensations
}
\author{Raul Horvat}
\affiliation{Physics Division, Rudjer Bo\v skovi\' c Institute, 
Zagreb, Croatia}
\author{Peter Minkowski}
\affiliation{Institute for Theoretical Physics, University of Bern,
CH-3012 Bern, Switzerland}
\author{Josip Trampeti\'{c}}
\affiliation{Theoretical Physics Division, Rudjer Bo\v skovi\' c Institute, 
Zagreb, Croatia}

\date{\today}

\begin{abstract}
In this paper we discuss light neutrino 
dipole moments, computed in the neutrino-mass extended standard model (SM),
as a possible source for neutrino condensates 
which may cause cosmological constant observed today.

\end{abstract}


\maketitle

In this paper we propose light neutrino long range dipole-dipole forces arising
from dipole moments, computed in the neutrino-mass extended standard model (SM),
as a possible source for neutrino condensates. 
These condensates will cause acceleration through the associated cosmological constant 
provided the vacuum pressure is dominant and negative \cite{O'Connell:2007xy}.
The computation depends on the nature of the neutrinos,
however, we first discuss the consequences of the neutrino-photon interaction 
with characteristic electromagnetic properties of Majorana neutrinos: 
the transition dipole moments 
\cite{FY,Raidal,Lee:1977tib,Shrock:1982sc,Heusch:1993qu}. 
These minuscule transition dipole moments are sensitive probes 
of fluctuations at scales as small as 
$10^{-35}\; {\rm cm}$ \cite{Duplancic:2003zw}, as seen through electromagnetic
interactions at long range. This can also shed more light on 
the expansion of the universe and the cosmological 
constant problem \cite{O'Connell:2007xy}.

The transition matrix elements relevant 
for $\nu_i \longrightarrow \nu_j;\;i\not= j$ 
in neutrino mass extended standard model for Majorana neutrinos are given in
\cite{FY,Raidal,Lee:1977tib,Shrock:1982sc,Heusch:1993qu}.
The photon-neutrino effective vertex is basicaly
determined from the $\nu_i \longrightarrow \nu_j\,\,\gamma$ transition, 
which is generated through electroweak processes that
arise from one-loop diagrams via the exchange of $\ell=e,\mu,\tau$
leptons and weak bosons, and is given by 
\begin{eqnarray}
J_{\mu}^{\rm eff}\epsilon^{\mu}(q)&=&
\left\{ F_1(q^2) {\bar \nu_j}(p')_L\,(\gamma_{\mu}q^2-q_{\mu}{\not \!q})\,\nu_i(p)_L 
\right. \nonumber\\
&-&\left.
{\rm i}F_2(q^2)\left[ m_{\nu_j}{\bar \nu_j}(p')_R\,\sigma_{\mu\nu}\,q^{\nu}\nu_i(p)_L 
 \right.\right. \nonumber\\
&+&\left.\left. 
m_{\nu_i}{\bar \nu_j}(p')_L\,\sigma_{\mu\nu}\,q^{\nu}\nu_i(p)_R\right]\right\}
\epsilon^{\mu}(q).
\label{1}
\end{eqnarray}
The above effective interaction is invariant under 
electromagnetic gauge transformations.
The first term in (\ref{1}) vanishes for 
real photon due to the electromagnetic gauge condition. 

The general decomposition of the $F_2$ term of 
the transition matrix element $T[A,B]$ obtained from (\ref{1}),
leads to the well known expression for the electric and magnetic dipole moments 
\begin{eqnarray}
d^{\rm el}_{ji}&=&\hspace{-2mm}\frac{-e}{M^{*2}}
\left( m_{\nu_i}-m_{\nu_j}\right)
\hspace{-3mm}\sum_{k=e,\mu,\tau}\hspace{-3mm}{\rm U}^{\dagger}_{jk}{\rm U}^{}_{ki}
F_2(\frac{m^2_{\ell_k}}{m^2_W}),
\label{6} \\
\mu_{ji}&=&\hspace{-2mm}\frac{-e}{M^{*2}}
\left( m_{\nu_i}+m_{\nu_j}\right)
\hspace{-3mm}\sum_{k=e,\mu,\tau}\hspace{-3mm}{\rm U}^{\dagger}_{jk}{\rm U}^{}_{ki}
F_2(\frac{m^2_{\ell_k}}{m^2_W}),
\label{7}
\end{eqnarray}
where $i,j=1,2,3$ denotes neutrino species, and  
\begin{eqnarray}
F_2(\frac{m^2_{\ell_k}}{m^2_W})\simeq -\frac{3}{2}+\frac{3}{4}
\frac{m^2_{\ell_k}}{m^2_W},
\;\;\;\frac{m^2_{\ell_k}}{m^2_W}\ll 1,
\label{8}
\end{eqnarray}
was obtained after the loop integration. In Eqs. (\ref{6}) and (\ref{7}) 
$M^*=4\pi\,v=3.1$ TeV, where $v=(\sqrt 2\, G_F)^{-1/2}=246$ GeV 
represents the vacuum expectation value of the scalar Higgs field \cite{Duplancic:2003zw}. 

Note that in the case of a mass degenerate pair the
electric dipole moment vanishes, while the magnetic one is dominated
by the first term in (\ref{8}). In the case of off-diagonal transition moments, 
the first term in (\ref{8}) 
vanishes in the summation over leptons due to the orthogonality 
condition of the neutrino mixing matrix U \cite{Maki:1962mu} (GIM cancellation).

The mixing matrix U is governing 
the decomposition of a coherently
produced left-handed neutrino $\widetilde{\nu}_{L,\ell}$ 
associated with charged-lepton-flavor $\ell = e, \mu, \tau$ into
the mass eigenstates $\nu_{L,i}$:
\begin{eqnarray}
|\widetilde{\nu}_{L,\ell};\,\vec p\,\rangle =
\sum_i {\rm U}_{\ell i} |\nu_{L,i};\,\vec p ,m_i\,\rangle ,
\label{9}
\end{eqnarray}

The characterizing feature of Majorana neutrinos
(i.e. 4-component notation the Hermitian, neutrino-flavor antisymmetric, 
electric and magnetic dipole operators), 
i.e.\ fields that do not distinguish particle 
from anti-particle ($\psi_i=\psi_i^c$), producing
a transition matrix element T[$A,B$] which is a complex antisymmetric 
quantity in lepton-flavor space:
\begin{eqnarray}
{\rm T_{ji}}&=&-{\rm i}\epsilon^{\mu}{\bar\nu_j}
\left[(A_{ji}-A_{ij})-(B_{ji}-B_{ij})\gamma_5\right]\sigma_{\mu\nu}q^{\nu}\nu_i
\nonumber\\
&=&-{\rm i}\epsilon^{\mu}{\bar\nu_j}
\left[2{\rm i}\,{\Im}A_{ji}-2{\rm \Re}B_{ji}\gamma_5\right]\sigma_{\mu\nu}q^{\nu}\nu_i\,,
\label{15}
\end{eqnarray}
i.e. antisymmetric with respect to neutrino mass eigenstates.
From this equation it is explicitly clear that for $i=j$, 
$d^{\rm el}_{\nu_i}={\mu}_{\nu_i}=0$.
Also, considering transition moments, only one of 
two terms in (\ref{15}) is 
non-vanishing if the interaction respects CP invariance: 
The first term vanishes if 
the relative CP of $\nu_i$ and $\nu_j$ is even, and 
the second term vanishes if it is odd \cite{Shrock:1982sc}.
Dipole moments describing the transition 
from Majorana neutrino mass eigenstate-flavor 
$\nu_j$ to $\nu_i$ in the mass extended standard model are:
\begin{eqnarray}
\hspace{-2mm}d^{\rm el}_{{\nu_j}{\nu_i}}\hspace{-1.5mm}&=&
\hspace{-1.5mm}\frac{3\,\rm i\,e}{2M^{*2}} 
\left( m_{\nu_i}-m_{\nu_j}\right)
\hspace{-2mm}\sum_{k=e,\mu,\tau}\frac{m^2_{\ell_k}}{m^2_W}
\;{\Re}({\rm U}^{\dagger}_{jk}{\rm U}^{}_{ki}),
\label{16}\\
\hspace{-2mm}{\mu}_{{\nu_j}{\nu_i}}\hspace{-1.5mm}&=&
\hspace{-1.5mm}\frac{3\,\rm i\,e}{2M^{*2}} 
\left( m_{\nu_i}+m_{\nu_j}\right)
\hspace{-2mm}\sum_{k=e,\mu,\tau}\frac{m^2_{\ell_k}}{m^2_W} 
\;{\Im}({\rm U}^{\dagger}_{jk}{\rm U}^{}_{ki})\, ,
\label{17}
\end{eqnarray}
where the neutrino-flavor mixing matrix 
U is approximatively unitary, i.e  
it is necessarily of the following form \cite{Duplancic:2003zw}
\begin{eqnarray}
\sum_{i=1}^3 {\rm U}^{\dagger}_{jk}{\rm U}^{}_{ki} = 
{\delta}_{ji} - \varepsilon_{ji},
\label{18}
\end{eqnarray}
where $\varepsilon$ is a hermitian nonnegative matrix 
(i.e. with all eigenvalues nonnegative) and 
\begin{eqnarray}
|\varepsilon|= \sqrt{{\rm Tr}\;\varepsilon^2} &=& 
{\cal O} \; (m_{\nu_{\rm light}}/m_{\nu_{\rm heavy}})
\sim 10^{-22} \;\, {\rm to}\;\, 10^{-21}.
\nonumber\\ 
\label{19}
\end{eqnarray}
It is important to note that the first term ${\delta}_{ji}$ from (\ref{18})
in our case does not contribute, and that 
the case $|\varepsilon|=0$ is excluded by the very existence 
of oscillation effects.

The transition dipole moments in general receive very small contributions 
because of the smallness of the neutrino mass, 
$|m_{\nu}|\simeq 10^{-2}$ eV \cite{nobel}. 
The largest contribution among them is proportional to ${\Re}$ and ${\Im}$ parts
of ${\rm U}^{\dagger}_{3\tau}{\rm U}^{}_{\tau 2}$, 
which corresponds to the $2 \to 3$ transition.
For the sum and difference of neutrino masses we 
assume hierarchical structure and take 
$|m_3 + m_2| \simeq |m_3 - m_2| \simeq |\Delta m_{32}^2|^{1/2} 
= 0.05$ eV \cite{nobel}. 
For the mixing matrix elements \cite{Maki:1962mu} we set 
$|{\Re}({\rm U}^{\dagger}_{3\tau}{\rm U}^{}_{\tau 2})| 
\simeq|{\Im}({\rm U}^{\dagger}_{3\tau}{\rm U}^{}_{\tau 2})| \le 0.5$.

The electric and magnetic transition dipole moments of neutrinos 
$d^{\rm el}_{\nu_2\nu_3}$ and $\mu_{\nu_2\nu_3}$ are then denoted 
as $\left(d^{\rm el}_{\rm mag}\right)_{23}$  
and are given by
\begin{eqnarray}
\left|\left(d^{\rm el}_{\rm mag}\right)_{23}\right| &=&
 \;\frac{3e}{2M^{*2}}\;
\frac{m^2_{\tau}}{m^2_W}\sqrt{|\Delta m_{32}^2|}
{|{\Re}({\rm U}^{\dagger}_{3\tau}{\rm U}^{}_{\tau 2})|\choose |
{\Im}({\rm U}^{\dagger}_{3\tau}{\rm U}^{}_{\tau 2})|},
\nonumber \\
&\stackrel{<}{\sim}& 2.03 \times 10^{-30} [\rm e/eV] = 0.38 
\times 10^{-34} \,[\rm e\;cm],
\nonumber \\
&=& 2.07\times 10^{-24}\, \mu_B.
\label{dM}
\end{eqnarray}

Note that neutrino mass extended standard model, as a consequence of loops (\ref{8}),
produces four orders of magnitude higher
moments for a Dirac neutrino versus Majorana neutrino (\ref{dM}), due to an 
$(m^2_{\ell}/m^2_W)$-suppression of Majorana moments 
relative to the Dirac ones \cite{foot}.
  
Also note that electric transition dipole moments of light neutrinos 
are smaller than the ones of the d-quark. 
This is {\it the} order of magnitude of light neutrino 
transition dipole moments underlying the see--saw mechanism \cite{gell}. 
It is by orders of magnitude smaller than in lepton flavor unprotected SUSY models.
See properties of neutrinos with respect to models which contain
flavor mixing, the mass spectrum, dipole moments, electroweak radius, ect.
including additional contributions arising from SUSY GUT's, 
extra dimensions, non-commutativity of space-time, etc., in
\cite{FY,Fritzsch:1998xs} and refs quoted therein.
Of course rigorously established experimental bounds on 
the dipole moments of neutrinos are 
by orders of magnitude weaker than implied by our hypotheses (\ref{dM}).
The properties of astrophysical neutrinos can be found in
the following references \cite{FY,Fukugita:1987uy}.
                
Up to this point our presentation is fully relativistic but valid only 
for not too large momenta as appropriate for the long range 
approximation adopted.

The non-relativistic components of electric and magnetic fields,
whose coefficients are our electric and magnetic dipole moments, are
\begin{eqnarray}
E_j(\vec{d}|_0)
&=& \frac{1}{4\pi} \,d_{k} \,\partial_{k} \, \partial_{j}\,\frac{1}{r}\,,
\label{eq:1.1}\\
B_j(\vec{\mu}|_0) 
& = & \frac{1}{4\pi} \, \mu_{k} 
\,[\partial_{k} \, \partial_{j}
\, - \, \delta_{kj} \, \Delta] \,\frac{1}{r}\,,
\label{eq:1.2}
\end{eqnarray}
For dipole "$\vec{d}$" at position "$0$" determined with position vector 
$\vec{x}_{0}$ we have the following fields at point $\vec{x}$:
\begin{eqnarray}
\vec{E}(\vec{d}|_0) 
& =&
\frac{1}{4\pi}
\left(3 \vec{e} (\vec{e}\vec{d})  -  \vec{d}\right)
\frac{1}{r^3}
 -  \frac{1}{3}  \vec{d}  \delta^{3}(\vec{x}  -  \vec{x}_{0})\, ,
\label{eq:2.1}\\
\vec{B}(\vec{\mu}|_0) 
& =&
\frac{1}{4\pi} 
\Big(3\vec{e} (\vec{e}\vec{\mu})  -  \vec{\mu}\Big)
\frac{1}{r^3}
 +  \frac{2}{3} \vec{\mu}  \delta^{3}(\vec{x}  -  \vec{x}_{0})\,. 
\label{eq:2.2}
\\
r&=&|\vec{x} \, - \,\vec{x}_{0}|,\, 
\vec{e}= (\vec{x} \, - \, \vec{x}_{0}) \, / r,\,
\partial_{n}  =  \partial/\partial_{x_{n}}\,.
\nonumber
\end{eqnarray}

For neutrinos in the non-relativistic equal dipole-dipole approximations they are of 
the form represented by hermitian operators whose matrix elements 
are given in Eqs (\ref{6},\ref{7},\ref{15},\ref{16},\ref{17}). 

Restricting to equal dipole-dipole interactions only in the case of transition 
$1\,\rightarrow\,2$ we define relative distance vector as 
$\vec{e}= (\vec{x}  - \vec{x}^{\prime} ) / r$ where $\vec{x}$ and $\vec{x}^{\prime}$
are position vectors of dipole $1$ and $2$ respectively, 
and obtain well known dipole-dipole potential
\begin{eqnarray}
V(d \,, d^\prime) \, 
&=&- \frac{1}{4\pi} \,\left(3 (\vec{d} \, \vec{e})
(\vec{d}^{\prime} \, \vec{e}) \, - \, (\vec{d} \, \vec{d}^{\prime} )\right)\,
\frac{1}{r^3} 
\nonumber\\
&& \phantom{\frac{1}{4}}+\frac{1}{3} (\vec{d} \, \vec{d}^{\prime}) \,  
\delta^{3}\,(\vec{x}^{\prime} \, - \, \vec{x}), 
\label{eq:3.1}\\
V(\mu \,, \mu^{\prime}) \, 
&=&- \frac{1}{4\pi} \Big(3 (\vec{\mu} \, \vec{e})
(\vec{\mu}^{\prime} \, \vec{e}) \, - \, (\vec{\mu} \, \vec{\mu}^{\prime} )\Big)\,
\frac{1}{r^3} 
\nonumber\\
&&\phantom{\frac{1}{4}}- \frac{2}{3} (\vec{\mu} \, \vec{\mu}^{\prime}) \,  
\delta^{3}\,(\vec{x}^{\prime} \, - \, \vec{x}). 
\label{eq:3.2}
\end{eqnarray}
The discussed above dipole moments give rise to electric and magnetic 
long range dipole-dipole forces, which
are the only ones in the non-relativistic setting.
Hence only the nonlocal terms in the potentials 
$V \,(d \,, d^{\prime} \ ) \,, V \, (\mu \,, \mu^{\prime})$
are of concern to us here.


Note that by introducing 
the gravitational potential for any neutrino pair 
\begin{eqnarray}
     V_{\rm gravity} 
\,=\, -G_N \,\delta_{j_{1}i_{1}}\,\delta_{j_{2}i_{2}} 
\frac{m_{\nu_{i_{1}}}\,m_{\nu_{i_{2}}}}{|\;r\;|},\;\;\;\; i_{1} < i_{2}\,.
\label{20}
\end{eqnarray}
and equating the generic absolute values of gravitational and dipole-dipole
potentials, (\ref{eq:3.2}) and (\ref{20}), at $r=R\not=0$, 
together with Eq. (\ref{dM}), we obtain the interesting characteristic distance
\begin{eqnarray}
R&=&\sqrt{\frac{\alpha_{\rm em}}{m_{\nu_{i_{2}}}\,m_{\nu_{i_{3}}}}}
\left|\frac{\left(d^{\rm el}_{\rm mag}\right)_{23}}{e}M_{Pl}\right|\,,
\nonumber\\
&=&\sqrt{\frac{\alpha_{\rm em}}{500}}\times 0.38\times 10^{-34}\;
\left(\frac{1\rm cm}{L_{Pl}}\right)\times 0.0197\,[\rm cm]\,,
\nonumber\\
&=& 1.77\times 10^{-6}\,[\rm cm]\,,
\label{21}
\end{eqnarray}
where the above unique long-range potentials are comparable.

We assume that light neutrino condensates, 
due to neutrino transition dipole moments interaction energy, 
are also responsible for formation of dark energy. 
To estimate the dark energy density due to $\nu$-dipole potentials, $\rho^{\nu}_{DED}$,  
we first find the absolute value of the caracteristic energy due to 
dipole-dipole interaction ${<\epsilon_{\nu}>}_{vac}$: 
\begin{eqnarray}
{<\epsilon_{\nu}>}_{vac}&\simeq&\frac{|\int d^3r\,V|}{v}
=\frac{1}{v}\Big|\frac{2}{3}\sum_{i,j=1}^{N_{\nu}}\vec{\mu}_i\vec{\mu}_j\Big|\,,
\label{22a}
\end{eqnarray}
where $v$ is an intrinsic volume and $N_{\nu}$ is number neutrino pairs. 
Next we define $|\mu|^2$ as caracteristic measure of quadratic dipole strenght:
\begin{eqnarray}
|\mu|^2 &=& \Big|\frac{2}{3}\sum_{i,j=1}^{N_{\nu}}\vec{\mu}_i\vec{\mu}_j\Big|\,,
\label{22b}
\end{eqnarray}
and then the dark energy density due to $\nu$-dipoles is  
\begin{eqnarray}  
\rho^{\nu}_{DED}& = & \frac{{<\epsilon_{\nu}>}_{vac}}{v} 
=\Big(\frac{|\mu|}{v}\Big)^2 \, .
\label{22}
\end{eqnarray}
This is maximal for the case  
for $\mu \times \mu^{\prime} = \mu_{21} \times \mu_{12} 
= - ( \mu_{12} )^2 = |\mu_{12} |^2 \simeq |\mu|^2$, etc. 
Namely in the two neutrino 
channel (both spins, i.e. $\nu$ and/or $\bar\nu$)
the dipole-dipole interactions do not change the total energy 
$\sim\;m_1+m_2$, provided that the pair is composed of two different mass-flavors,
i.e. $1 \not= 2$ (in the s-channel). 
Antisymmetric type of interactions just changes the flavor ordering
$\nu_{1 , m_{1}} \nu_{2 , m_{2}} \rightarrow  \nu_{2 , m_{2}} \nu_{1 , m_{1}}$, 
(i.e. $m_{1} \leftarrow\rightarrow m_{2}$ at fixed $1,2$).
This gives the overall contribution, for $d_{ij}\rightarrow d_{12}$
with $ij$ mass-eigenstate-flavors, which is, for example,
$d \times d^{\prime} = d_{21} \times d_{12} = - ( d_{12} )^2 = |d_{12} |^2$
because of the factor ${\rm i}^2$ coming from $-( d_{12} )^2$.
Thus the attraction or repulsion is within one mass-pair-channel and thus 
fully active without changing the mass
of the pair provided of course 
the {\it mass-flavors in the pair are distinct}.

In this way identifing, by hypotesis, (\ref{22}) 
with the measured dark energy density today $\rho_{DED}$ we have found:
\begin{eqnarray}
v&=&\Big(\frac{|\mu|^2}{\rho_{DED}}\Big)^{1/2}=\frac{4\pi}{3}R^3_{\nu}\,, 
\nonumber\\ 
R_{\nu}&=&\Big(\frac{9}{16\pi^2} \frac{|\mu|^2}{\rho_{DED}}\Big)^{1/6}\,,
\label{23}
\end{eqnarray}
where $R_{\nu}$ is linear size of intrinsic volume $v$.

If we choose for $|\mu|$ the value of the dipole moments in Eq. (\ref{dM})
and from observation
$\rho_{DED}=(2.3\; \rm meV)^4\times \frac{h^2}{0.5}$, with $h=0.73$ being 
present day normalized Hubble constant \cite{Yao} we obtain:
\begin{equation}
R_{\nu}=0.84\times 10^{-13} \rm cm\,\simeq\, (200\,\rm MeV)^{-1}\,.
\label{24}
\end{equation}

Note that $R_{\nu}^{-1} \simeq 200$ MeV relates intrinsic volume $v$ to a
cosmological period corresponding to $T\sim R_{\nu}^{-1}$
which represents a distant past of cosmological evolution.

From (\ref{21}) and (\ref{24}) it follows $R_{\nu} \ll R$
which is consistent with the dipole moment interaction dominating 
gravitational ones. 

The elementary 4-neutrino interaction energy density is obviously very small,
but it has a collective $(number\;of\;neutrinos)^{2}$ growth.
In addition it definitely will have, for arbitrary moments otherwise,
an attractive sub-channel, depending on neutrino spins.
The attraction will generate condensation phenomena,
i.e. {\it neutrino-condensates}, by the Fermi-criterion, 
and since gravity is always attractive those two facts together lead to 
neutrino condensation phenomena relative to a free fermion gas.

Here we only consider condensates giving rise to a cosmological term
or equivalently to vacuum energy-density.
Assuming nonvanishing neutrino condensates due to dipole and gravitational long range
interactions giving rise to a cosmological term, 
(and/or to vacuum energy density and pressure),
i.e. not canceled by a readjustment 
of gravitational effects, it follows that these condensates are 
a specific source of dark energy.
The condensate will correspond eventually to some 'vacuum-energy 
density' and may not be canceled as all other larger condensates, 
e.g. of QCD, electroweak \cite{Brodsky:2008xu}, etc.

The condensate will then alter the neutrino energy-momentum dependence
as compared with free massive neutrino motion and thus the mean energy density 
in neutrinos will be larger for a given thermal ensemble and the same 
temperature. This temperature is approximately $2^\circ K$ today in 'the universe' 
and it corresponds to $\nu$-number density $n_{\nu_{0}}$, i.e. 
$n_{\nu_{0}} \simeq$ 300 free neutrinos per $\rm cm^3$ at present.



From known cosmological parameters we have  
dark energy density today, 
while energy density of free light neutrinos  
(above vacuum), at temperature $2^\circ K$ 
and assuming neutrino mean mass $m_\nu \simeq 20$ meV, is
$\rho_{ED}=(0.4634\; \rm meV)^4$. 
Ratio of those two facts
\begin{eqnarray}
\frac{(dark\, energy\, density)}
{(\nu-number \,density)\,\times\, (\nu-mean\, mass)}\,,
\nonumber
\end{eqnarray}
produces an interesting experimental number:
\begin{eqnarray}
\frac{\rho_{DED}}{\rho_{ED}}
=
(\frac{2.3369}{0.4634})^4=5.043^4 \simeq 647.
\label{25}
\end{eqnarray} 
Neutrino mean mass of 20 meV was used due to 
the assumption of normal neutrino family hierarchy. Of course this
number is larger in the case of inverted hierarchy.

If our analysis can overcome the factor 647 and furthermore, 
since we are comparing two very different types of energy densities, 
this could be transfered to neutrino condensates. 

The experimental ratio in Eq. (\ref{25}) has no direct bearing on the
size of the neutrino condensates, which represent vacuum energy density.
It is used  only here in order to emphasize that we cannot exclude 
the possibility that  the sum of neutrino condensates equals the 
observed dark energy density $\rho_{DED}$
in value $(2.34 \;\rm meV )^{4}$ and sign (positive),
causing acceleration of the universe expansion today 
(and tomorrow) and beeing de Sitter-like.
Our entire approach also ilustrates the sign of dark energy density which is
inconsistent with stability in the framework of local field theory
in uncurved space-time.

This could be related to another inconsistency arising from large, but finite,
lifetimes of not only light neutrions and 'baryons'. 
Our estimate of unstable neutrino lifetime from the decay rate
in the neutrino-mass extended standard model (SM)
\begin{eqnarray}
\Gamma(\nu_h \to \nu_{\ell}\,\gamma) 
&=& \frac{m_{\nu_h}^5}{16\pi}\,
\Big(\frac{G_F}{\sqrt2} \frac{e}{4\pi^2} \,{\rm U}^{\dagger}{\rm U}\,F_2\Big)^2 
\nonumber\\
&\simeq& 1.6\times 10^{-63}\,\rm meV\,,
\label{26}
\end{eqnarray}
gives 
\begin{equation}
\tau_{\nu_h}
 \simeq 4\times 10^{51}\,\rm s\,.
\label{27}
\end{equation}
This value was obtained from $\nu$-dipole moment interaction (\ref{1})-(\ref{dM})
with neutrino mass: $m_{\nu_h}=50$ meV.


It is interesting to notice that due to the sign of (\ref{22a}-\ref{22}), the
total energy density, $\rho$, of relic neutrinos,
\begin{equation}
\rho = m_{\nu } n_{\nu } - \rho_{DED}^{\nu}(=n_{\nu }^2 |\mu|^2)\,,
\label{28}
\end{equation}
may have an extremum during the cosmological evolution. 
Namely $n_{\nu }=n_{{\nu}_0}\,a^{-3}$, with $a$ being the scale factor of the universe.
This, in turn, may
entail consequences for an  accelerating phase of the expansion of the
universe, since near extremum $a_{ext}$, the EOS for relic neutrino gas,
\begin{eqnarray}
w_{\nu } +1 = -\frac{1}{3}\frac{\partial}{\partial({\rm log}\,a)}({\rm log}\,\rho)
=-\frac{1}{3}\frac{a}{\rho}\frac{d\rho}{da}\,, 
\label{ext} 
\end{eqnarray}
switches to $\approx -1$\,. Indeed, from
(\ref{28}) we find
\begin{equation}
a_{ext} = \big(\frac{2n_{{\nu }_0} |\mu|^2}{m_{\nu }}\big)^{1/3} \;.
\label{29}
\end{equation}
If we are to explore effects for a late-time accelerating phase, then we
should set $a_{ext} \sim 1$. However, even for magnetic moments as large as
$10^{-10} \mu_B $, the neutrino mass would be hopelessly tiny to induce any
observable effect on present acceleration of the universe.

As a way out of the above inconsistency one can recall a recent model proposed by Fardon,
Nelson and Weiner (FNW) \cite{FNW} and developed later by Kaplan, Nelson and 
Weiner \cite{KNW}, and Peccei \cite{RP}, in which relic neutrinos are tied together
with the sector of `standard' dark energy (represented by a canonically
normalized scalar field). The model is very appealing with regard
to the `cosmic coincidence problem' \cite{PJ}, since from the known behavior
of dark matter, ordinary matter and radiation one finds that any
reasonable tracking of these components by dark energy
always goes at the  expense
of the late time transition of its equation of state, thus creating a new
problem called the "why now?" problem. On the other hand, if relic neutrinos
can be kept tightly coupled  to the original dark energy fluid for most of
the history of the universe, the near coincidence at present, 
$\rho_{\Lambda}\sim \rho_{\nu }$, 
will cease  to be perceptive as a coincidence at all.
This was possible if the mass of the neutrino was promoted to a dynamical
quantity, being a function of the acceleron field (canonically normalized
scalar field similar to quintessence). The main feature of the
scenario \cite{FNW} is that although the number density of neutrinos dilutes
canonically $(\sim a^{-3})$, the masses of  neutrinos change almost
inversely $(\sim a^{-3w })$, thereby promoting their energy density to
an almost undilutable substance. Hence relic neutrinos become tightly
coupled to the original dark energy fluid.        

In addition, by applying the FNW scenario to our model, in which the energy
density for relic neutrinos is supplemented with a term due to nonvanishing
electro-magnetic moments, we can draw some conclusions about intrinsic
properties of neutrinos if also $|\mu|$ is considered as a dynamical
field (some function of $m_{\nu })$. In this case one can show that in the
FNW scenario the EOS for the coupled dark energy fluid obeys
\begin{equation}
w + 1 = \frac{m_{\nu }n_{\nu } -2 m_{\nu }^2 |\mu|^2}{\rho_{totaldark}}.
\label{30}
\end{equation}
Since the neutrino contribution gives only a small fraction of the total energy
density, we have $w \simeq -1$, in accordance with what cosmological data
imply. Also, the data imply very slow variation of $w$ with $a$, which,
taken in a literal sense, means that both terms in the numerator of (\ref{30})
should scale as $\rho_{totaldark} \sim a^{-3(1 +w)}$. This entails, 
$m_{\nu }\sim a^{-3w }$, $|\mu|^2 \sim a^{-3(1 - w)}$. It is interesting to
note that although the scaling of $m_{\nu }$ and $|\mu|$ with $a$ 
are formally different, they become the same in the limit $w \rightarrow -1$. 
This complies with the prediction of the minimally extended SM
$|\mu|_R \sim |\mu|$, where to each generation of fermions of the SM a
right-handed neutrino field is added, in
contrast with more complicated models where the neutrino magnetic moment is
disentangled from the neutrino mass.
  
In conclusion, we have considered the cosmological consequences of
long-range interactions in a non-relativistic setting and arising from
various electromagnetic form factors of a neutrino. We have emphasised 
the possibility that the responsible interaction  itself has an attractive
channel, leading neutrino condensation phenomena to occur. This would
entail a sort of dark energy, responsible for the late-time acceleration in
the expansion of the universe. In addition, the energy density due to neutrino
electromagnetic moments, when superimposed  on the 
standard contribution of a
neutrino background, may be responsible for acceleration phases during the
history of the universe. 
When implemented in 
a recently suggested dark energy scenario with mass varying neutrinos, the
electromagnetic neutrino interaction may also shed some light on intrinsic
neutrino properties. 

\vspace{2.5cm}

The work of R.H. and J.T. is supported by the Croatian Ministry of Science, 
Education and Sport under Contract 
No. 0980982930-2872 and No. 0980982930-2900.

\end{document}